\documentclass[aps,prb,twocolumn,showpacs,amsmath,amssymb,unsortedaddress]{revtex4-1}

\pdfoutput=1
\usepackage[english]{babel}
\usepackage[utf8]{inputenc}
\usepackage{amsmath} 
\usepackage{mathtools}
\usepackage{mathrsfs}
\usepackage{graphicx}
\usepackage[colorinlistoftodos]{todonotes}
\usepackage{afterpage}

\usepackage{dcolumn}% Align table columns on decimal point
\usepackage{bm}% bold math
\usepackage{url}
\usepackage[colorlinks=true,linkcolor=blue,citecolor=blue,urlcolor=blue]{hyperref}
\usepackage[normalem]{ulem}

\usepackage{color}

\def\zRevisions#1{\textcolor{purple}{#1}}

\begin{document}

%\title{Revisiting Debye-Peirels lattice thermal conductivities: Lower dimensionality, anomalous dispersion, and quasi-amorphism}
\title{A generalized Debye-Peierls/Allen-Feldman model for the lattice thermal conductivity of low dimensional and disordered materials }

\author{Taishan Zhu}
\affiliation
{Department of Mechanical Science \& Engineering, University of Illinois at Urbana-Champaign, IL 61820} % 
\author{Elif Ertekin}
\affiliation
{Department of Mechanical Science \& Engineering, University of Illinois at Urbana-Champaign, IL 61820}
\affiliation
{International Institute for Carbon Neutral Energy Research (WPI-I$^2$CNER), Kyushu University, 744 Moto-oka, Nishi-ku, Fukuoka 819-0395, Japan}
% % \date{} %\today

\begin{abstract}
We present a generalized model to describe the lattice thermal conductivity of low-dimensional (low-D) and  disordered systems. 
The model is a straightforward generalization of the Debye-Peierls and Allen-Feldman schemes to arbitrary dimensions, accounting for low-D effects such as differences in dispersion, density of states, and scattering. 
Similar in spirit to the Allen-Feldman approach, heat carriers are categorized according to their transporting capacity as propagons, diffusons, and locons. 
The results of the generalized model are compared to experimental results when available, and equilibrium molecular dynamics simulations otherwise. 
The results are in very good agreement with our analysis of phonon localization in disordered low-D systems, such as amorphous graphene and glassy diamond nanothreads. 
Several unique aspects of thermal transport in low-D and disordered systems, such as milder suppression of thermal conductivity and negligble diffuson contributions, are captured by the approach.
\end{abstract}

\maketitle

%\begin{description}
%\item[KEYWORDS]: Carbon allotrope, diamond nanotreads, carbon nanotubes, Stone-Wales defects, phonon, thermal properties
%\end{description}

% \pacs{05.60.-k, 63.20.-e, 63.22.-m, 66.70.-f, 68.65.Cd}
% PACS numbers(s): 05.60.-k, 63.20.-e, 63.22.-m, 66.70.-f, 68.65.Cd
% \tableofcontents

\newpage

%%%%%%%%%%%%%%%%%%%%%%%%%%%%%%%%%%%%%%%%
%%%%%%%%%%%%%%%%%%%%%%%%%%%%%%%%%%%%%%%%
%%%%%%%%%%%% MANUSCRIPT  %%%%%%%%%%%%%%%
%%%%%%%%%%%%%%%%%%%%%%%%%%%%%%%%%%%%%%%%
%%%%%%%%%%%%%%%%%%%%%%%%%%%%%%%%%%%%%%%%

\section{Introduction}

As thermal science expands into the realm of low-dimensional (low-D) materials  \cite{ Shi2012ThermalSystems, Balandin2011ThermalMaterials, Marconnet2013ThermalMaterials, Pereira2013ThermalCarbon, Luo2013NanoscaleExperiment,Cahill2003NanoscaleTransport,Cahill2014Nanoscale20032012, Chen2005NanoscalePhotons}, a variety of intriguing effects are being revealed \cite{Lepri2003ThermalLattices,Liu2014AnomalousDiffusion,Cahill2003NanoscaleTransport,Cahill2014Nanoscale20032012,  Chen2005NanoscalePhotons}.  
The unique physics of thermal transport in low-D has inspired many potential applications in the real world\cite{Pop2010EnergyDevices, Dresselhaus2005NewResearch,Marino2011Atomic-scaleTheory}. 
This physics is pushed to its limits when materials are genuinely atomically thin, such as two-dimensional graphene\cite{Ferrari2006RamanLayers} and one-dimensional carbon nanotubes (CNTs)\cite{Shi2012ThermalSystems,Marconnet2013ThermalMaterials, Pereira2013ThermalCarbon, Pettes2011InfluenceGraphene}.  
Like their three-dimensional analogs, low-D materials can also exhibit structural disorder.  
For instance, nanoporous graphene \cite{Surwade2015WaterGraphene.} and glassy diamond nanothreads \cite{Fitzgibbons2014Benzene-derivedNanothreads, Roman2015MechanicalNanothreads} are recent examples of materials systems in which both disorder and low dimensionality are simultaneously present. 
Ample applications of these materials are in incubation \cite{Surwade2015WaterGraphene.,Fitzgibbons2014Benzene-derivedNanothreads}, including thermoelectrics \cite{Dresselhaus2005NewResearch} and thermal barrier coatings \cite{Marino2011Atomic-scaleTheory}.  
However, while the structural, electronic \cite{Kotakoski2011FromCarbon,VanTuan2012InsulatingMembrane}, and mechanical properties\cite{Surwade2015WaterGraphene.,Fitzgibbons2014Benzene-derivedNanothreads,Roman2015MechanicalNanothreads} of low-D and disordered materials have received comparably greater attention, their thermal properties are less established.  

%As disorder is known to dramatically affect phonon transport and the lattice thermal conductivity ($\kappa$), this work is concerned with understanding how disorder affect the thermal properties of low-D materials. 

When disorder is strong enough that phonon mean free paths become comparable to phonon wavelengths, the quasiparticle picture breaks down. 
A different approach is required to describe vibrational transport, and disorder models that address this regime have been established for 3D. 
For weakly disordered systems, perturbation theory is reasonably accurate \cite{Klemens1955TheImperfections,Garg2011RoleStudy}. 
Towards the  fully amorphous limit, models such as random-walk \cite{Cahill1989HeatGlasses},  Allen-Feldman (AF)\cite{Allen1999DiffusonsSi}, and two-level states (TL) \cite{Anderson1972AnomalousGlasses,Sheng1991HeatStructures} are available.   
The random-walk picture was initiated by Einstein and later extended by Cahill \& Pohl and provides an estimate of the minimum thermal conductivity, the so-called {\it amorphous limit}. \cite{Cahill1989HeatGlasses} 
In comparison to disordered 3D systems, several interesting questions arise regarding the thermal physics of disorder in low-D. 
On one hand the thermal conductivity $\kappa$ of low-D materials such as graphene and carbon nanotubes can be exceptionally large (suggested in some cases to \zRevisions{diverge with increasing system size} \cite{Xu2014Length,ChangPRL08}). 
On the other hand, the effects of disorder (or any other perturbation) are typically more pronounced in low-D. 

Our recent work has focused on localization analysis of vibrational modes and equilibrium molecular dynamics simulations  of $\kappa$ using two examples ofgeneralized model low-D, disordered materials~\cite{ZhuErtekinNL}: one dimensional (1D) glassy diamond nanothreads and two-dimensional (2D) amorphous graphene.  
Our equilibrium molecular dynamics simulations revealed that the suppression of $\kappa$ in both of these systems is small, in comparison to suppression commonly observed in 3D materials. 
In glassy nanothreads $\kappa$ drops by a factor of five in the presence of strong disorder, and only drops by 25\% in amorphous graphene. 
This is remarkably weak in comparison to 3D materials for which the suppression can be two to four orders of magnitude\cite{Cahill1989HeatGlasses}. 
Localization analysis of the modes suggest that the mild suppression arises from the resilience of transverse twist modes in the nanothreads and flexural modes in amorphous graphene. 
These modes appear to retain their wave-like character despite the structural disorder\cite{ZhuErtekinNL}.  
%On the other hand, the thermal conductivity of the disordered 1D diamond nanothread demonstrated a more typical suppression of $\kappa$, and a transition in the nature of the modes to include more longitudinal vibrations.  

In this work, we present a generalized model that describes vibrational transport in low-D and disordered materials. 
While state-of-the-art computational modeling of thermal transport in low-D and/or disordered materials is now possible and has revealed many insights\cite{Luo2013NanoscaleExperiment,Garg2011RoleStudy,Wang2007Length-dependentMeasurements, Yamamoto2006NonequilibriumNanotubes}, simplified approximate models that capture the physics without requiring the full solution can also be very useful. \cite{Allen2013ImprovedConductivity}  
Such models often give insight into essential underlying mechanisms and can quickly reproduce or predict trends.  
Our model describes the thermal conductivity $\kappa$ and its temperature ($T$) dependence in disordered, low-D materials. 
The results are in good agreement with experiment measurements of low-D systems that have been reported in the literature, or equilibrium molecular dynamics simulations of $\kappa$ for diamond nanothreads and amorphous graphene~\cite{ZhuErtekinNL}. 
This illustrates that when formulated properly simple models can reproduce trends even at these scales. 
The analysis of disorder presented here is specific to the case of 1D diamond nanothreads and 2D amorphous graphene, but the framework is general and can apply as well to other low-D, disordered systems as well.  

\section{Thermal Conductivity of Crystalline Materials in Arbitrary Dimensions}

For crystalline 3D materials, models of phonon thermal conductivity are well-established \cite{Peierls1955QuantumSolids,Callaway1959ModelTemperatures,Holland1963AnalysisConductivity,Klemens1994ThermalPlane,Allen2013ImprovedConductivity}. 
In \zRevisions{1929} Peierls formulated the lattice conductivity of bulk dielectric crystals in terms of the phonon Boltzmann transport equation \cite{dalitz1997selected, Peierls1955QuantumSolids}.  
Callaway, in 1959, introduced an approximate solution of the Peierls Boltzmann equation within the relaxation time approximation invoking a Debye description of solids that separately accounts for Normal and Umklapp scattering events \cite{Callaway1959ModelTemperatures}. 
This successfully reproduced the $\kappa$ {\it vs.} $T$ dependence of germanium for low temperatures. 
By further differentiating longitudinal acoustic (LA) and transverse acoustic (TA) phonons, Holland extended Callaway's model and achieved better high-temperature agreement \cite{Holland1963AnalysisConductivity}. 
These models, and several others that followed, have proven extremely useful for  understanding phonon transport in conventional 3D materials. 

We begin with an approach for crystalline materials reminiscent of the Callaway--Holland model but applicable in arbitrary dimensions. 
It accounts for variations of the phonon density of states and parabolic dispersions that can arise in low-D. 
This will be extended to amorphous or disordered systems in the next section. 
$\kappa_{\hat\imath}$, the thermal conductivity in direction $\hat\imath$, is given by a sum over contributions of phonon mode branches $m$ and phonon wave vectors $\vec{q}$ 
\begin{equation}
\kappa_{\hat\imath} = \sum_m \sum_{\vec{q}} \left( \vec{v} (\vec{q}_m) \right) \cdot \hat\imath)^2 \,  \tau(\vec{q}_m) \,  C_{ph}(\vec{q}_m)  \hspace{1em}, 
\label{eq:kappa} 
\end{equation}
where $\vec{v}(\vec{q}) = d\omega/ \, d\vec{q}$ is the group velocity, $\tau (\vec{q})$ is the mode-specific scattering time, and $C_{ph} (\vec{q})$ is the mode-specific heat capacity. 
For an isotropic solid with frequency-dependent phonon density of states $g(\omega)$, this becomes 
\begin{eqnarray}
\kappa & = & \sum_m \langle \mathrm{cos}^2 \theta \rangle \int v(\omega)^2 \, \tau(\omega,T) \, C_{ph}(\omega,T) \,  g(\omega) \, d\omega \nonumber \\
& = & \sum_m \frac{1}{d} \int v(\omega) \, \Lambda(\omega,T) \, C_{ph}(\omega,T) \,  g(\omega) \, d\omega \hspace{1em}, 
\label{eq:kappa2}
\end{eqnarray}
\zRevisions{where the sum over $\vec{q}$ in Eq. (\ref{eq:kappa}) has been converted into an integral over modal frequencies}, $\theta$ is the angle between wave vector $\vec{q}$ and a temperature gradient $\nabla T$, $d$ is the dimension, and the geometric factor $\langle \mathrm{cos}^2 \theta \rangle = 1/d$ arises from summing over modes propagating in all directions. 
The modal heat capacity is $C_{ph} = k_B x^2 e^x (e^x-1)^2$, where $x = \hbar\omega/k_BT$ with  Boltzmann constant $k_B$ and reduced Planck constant $\hbar$. 
In Eq. (\ref{eq:kappa2}), we give the expression for $\kappa$ both in terms of scattering rates $\tau(\omega,T)$, as well as in terms of modal mean free paths \zRevisions{$\Lambda(\omega,T) = v(\omega) \tau(\omega,T)$. 
For nanostructured systems, mean free paths may be more accessible and in the ballistic regime $\Lambda(\omega,T)$ can be set to a characteristic length scale or feature size. 
Alternatively, in the diffusive regime appropriate descriptions of  scattering times $\tau(\omega,T)$, such as those in Table I of Ref. [\onlinecite{Holland1963AnalysisConductivity}], can be used instead.}

\begin{table*}[h]
\caption{Density of states $g(\omega)$, group velocity $v$, and cutoff frequency $\omega_c$ for modes with linear and parabolic dispersion. Here, $s_d$ is the surface area of a $d$-dimensional sphere with unit radius: $s_d = (2, 2\pi, 4\pi)$ for $d = (1,2,3)$. $n$ denotes the average atomic spacing.}
\label{tab:summary}
\begin{tabular}{l c c c }
\hline
\hline
Dispersion & density of states $g(\omega)$  & group velocity $v$ $$ & cutoff frequency $\omega_c$ \\\hline 
$\omega = v q$ (linear) &  
\parbox{4cm}{ \begin{equation} \nonumber \frac{s_d \omega^{d-1}}{(2 \pi v)^d } \end{equation}} &  
$v$ & 
\parbox{4cm}{ \begin{equation} \nonumber 2 \pi v \left( \frac{nd}{s_d} \right)^{1/d} \end{equation}}  \\
$\omega = a q^2$ (parabolic) & 
\parbox{4cm}{ \begin{equation}  \nonumber
\frac{s_d}{2(2\pi)^d}a^{-\frac{d}{2}}\omega^{\frac{d}{2}-1}
\end{equation}} & 
$2 \sqrt{a\omega}$ & 
\parbox{4cm}{ \begin{equation} \nonumber
(2\pi)^2 a \left( \frac{nd}{s_d} \right)^{2/d}
\end{equation} }   \\
\hline \hline
\end{tabular}
\end{table*}

In 3D materials, due to translational lattice symmetry in all three directions the dispersion of the longitudinal and transverse acoustic modes is always concave, but this is not the case for low-D materials. 
For a 2D system such as graphene, in addition to the two in-plane branches (LA and TA), out-of-plane flexural modes (ZA) with parabolic, convex dispersion are present. 
The different dispersion arises from the governing wave equations. 
For longitudinal, transverse, and torsional branches, $\partial_t^2{\phi}=c^2 \partial_x^2{\phi}$ where the wave speed $c=\sqrt{M/\rho}$ and $M$ is the elastic modulus and $\rho$ the density. 
For flexural branches $\partial_t^2{\phi}=a^2 \partial_x^4{\phi}$ where $a^2=E I/\rho A_c$, and $EI,\rho, A_c$ are respectively the flexural rigidity, density, and cross-section. 
Due to the presence of these modes, the Debye model universally assumed in 3D can not be directly applied in low-D.  
In addition to different group velocities and density of states, the physics of scattering (and thus scattering times) may differ for parabolic modes. 
Table \ref{tab:summary} gives the density of states, group velocities, and cutoff frequencies applicable to linear and parabolic dispersion. 

For phonon branches \zRevisions{$m$} with linear dispersion  $m= \ell$ and parabolic disperion $m = p$, from Eq. (\ref{eq:kappa2}) and Table \ref{tab:summary} the corresponding contributions to $\kappa$ are 
\begin{widetext}
\begin{eqnarray}
\kappa_\ell & = & \frac{s_d}{d} \frac{k_B}{(2\pi)^d } \left( \frac{\zRevisions{k_B} T}{\hbar} \right)^d v^{1-d} \int_0^{x_c} \frac{\Lambda(x,T) x^{d+1} e^x}{(e^x-1)^2} \, dx \label{eq:einstein} \\
& = & \frac{s_d}{d} \frac{k_B}{(2\pi)^d } \left( \frac{k_B T}{\hbar} \right)^d v^{2-d} \int_0^{x_c} \frac{\tau(x,T) x^{d+1} e^x}{(e^x-1)^2} \, dx \hspace{1em}, \\
\kappa_p & = & \frac{s_d}{d} \frac{k_B}{(2\pi)^d } \left( \frac{k_B T}{\hbar} \right)^{\frac{d+1}{2}} a^{\frac{1-d}{2}} \int_0^{x_c} \frac{\Lambda(x,T) x^{\frac{d+3}{2}} e^x}{(e^x-1)^2} \, dx \label{br} \\
& = & \frac{2 s_d}{d} \frac{k_B}{(2\pi)^d } \left( \frac{k_B T}{\hbar} \right)^{\frac{d+2}{2}} a^{\frac{2-d}{2}} \int_0^{x_c} \frac{\tau(x,T) x^{\frac{d+4}{2}} e^x}{(e^x-1)^2} \, dx \hspace{1em}. 
\end{eqnarray}
\end{widetext}
The different scaling of $\kappa$ with $T$ exhibitted by linear {\it vs.} parabolic dispersion is one of physical distinctions that can arise in low-D.
In the expressions above, $s_d$ is the surface area of a $d$-dimensional sphere of unit radius,  and the dispersions for linear and parabolic modes are given by $\omega = vq$ and $\omega = aq^2$ respectively, see Table \ref{tab:summary}.

\section{Effects of Disorder} 

Next, we modify the approach to account for amorphous or disordered systems. 
For fully amorphous systems, we generalize the Cahill-Pohl 3D random-walk approach to arbitrary dimensions.  
To account for the disordered intermediate regime, we use an approach motivated by Allen and Feldman, in which phonon carriers are categorized according to their degree of localization and their mobility.  
This provides a simple framework to understand low-D phonon transport on materials ranging from crystalline to amorphous, and the results of the model are in good agreement with experimental results (when available) and our computational simulations (otherwise). 
We also make predictions for the scaling behavior of $\kappa$ {\it vs.} $T$ for low-D and disordered materials such as defective and amorphous graphene and disordered carbon nanothreads, which to our knowledge have not yet been measured or reported.

We start from Cahill's model for fully amorphous solids, which extended the original model proposed by Einstein \cite{Cahill1989HeatGlasses}.  
\zRevisions{In the original Einstein model, the thermal conductivity of a 3D amorphous solid is obtained from Eq. (\ref{eq:einstein}) \zRevisions{by setting the mean free path to the mean atomic spacing $\Lambda= n^{-1/3}$ where $n =$  (total number of atoms/total volume). }
The small mean free path reflects the localized nature of vibrational carriers in amorphous solids, which transport heat via short, diffusive ``random walk steps". 
Cahill's approach to amorphous solids allows for more delocalized vibrations (as suggested by Debye and Slack\cite{Slack1979TheCrystals}) by using a larger mean-free path $\Lambda = \lambda/2=\pi v/\omega$ equal to half the modal wavelength $\lambda$ for the random-walk step in Eq. (\ref{eq:einstein})}.  
Cahill's model yields satisfactory agreement with experiment for various 3D glassy materials. 
Substituting the random-walk step $\Lambda = \pi v/\omega$ into Eqs. (3) and (5), we can generalize Cahill's model to arbitrary dimensions for both linear and parabolic dispersion: 
\begin{widetext}
\begin{eqnarray} 
\label{eq:kCahill_d}
\kappa^{am}_\ell & = & \pi \frac{s_d}{d} \frac{k_B}{(2\pi)^d } \left( \frac{k_B T}{\hbar} \right)^{d-1} v^{2-d} \int_0^{x_c} \frac{ x^{d} e^x}{(e^x-1)^2} \, dx \hspace{1em}, 
\\
\kappa^{am}_p & = & \frac{s_d}{d} \frac{k_B}{(2\pi)^{d-1} } \left( \frac{\zRevisions{k_B} T}{\hbar} \right)^{\frac{d}{2}} a^{1-\frac{d}{2}} \int_0^{x_c} \frac{ x^{1+\frac{d}{2}} e^x}{(e^x-1)^2} \, dx \hspace{1em}. 
\end{eqnarray}  
\end{widetext}

\begin{figure*}[!hbtp]
\centering
\includegraphics[width=0.9\textwidth]{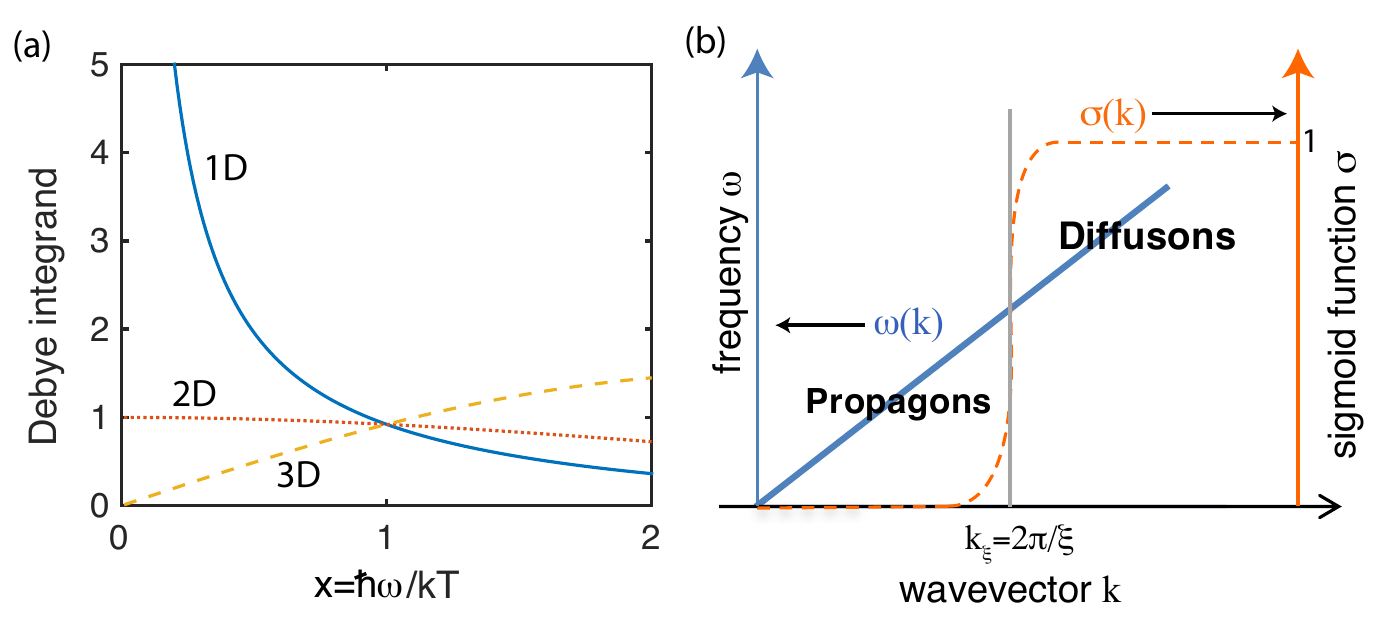}
\caption{\label{fig1} (a) The integrand of $\kappa_{\ell}^{am}$ from Eq. (\ref{eq:kCahill_d}), which converges for $d=2,3$, but diverges for $d=1$. (b) Schematic of the boundary $\sigma(k)$ demarcating propagons and diffusons. The Ioffe-Regel transition occurs around wavenumber $k = 2\pi/\xi$.}
\end{figure*}

Note that the original Cahill formula for linear modes is recovered for $d=3$. 
However, $\kappa^{am}_\ell$ derived above diverges for 1D glasses (the integrand is plotted in Figure \ref{fig1}(a)). 
The divergence can be traced back to the non-vanishing 1D density of states of long-wavelength phonons near the $\Gamma$ point. 
Such a divergence is unphysical and \zRevisions{is also related to the  large random walk step $\Lambda = \lambda/2=\pi v/\omega$ assigned to the modes in the low frequency limit $\omega \rightarrow 0$ in the Cahill approach}. 

Based on these considerations, we instead implement an approach based on Allen-Feldman (AF) theory \cite{Allen1999DiffusonsSi}, in which heat carriers -- so-called {\it vibrons} -- are catergorized according to their degree of localization.  
Vibrons are composed of extendons and locons, the former (typically low frequency modes) are spatially extended and the latter (typically high frequency modes) are localized.  
The boundary is called the mobility edge. 
Extendons contribute the most to the thermal conductivity, and they are further categorized as propagons and diffusons.  
\zRevisions{Propagons are the lowest-frequency members that transport heat in a manner reminiscent of typical phonons, while by contrast, diffusons remain spatially delocalized but  transport heat {\it via} diffusive random walk steps}. 
\zRevisions{The Ioffe-Regel boundary $\lambda = \xi$ represents the wavelength of the propagon/diffuson crossover and is an important parameter for obtaining an accurate and descriptive theory.
Here and hereafter $k_\xi = 2\pi/\xi$ denotes the wavenumber and $\omega_\xi$ the frequency corresponding to the crossover}.

\zRevisions{In our formalism, the boundary between propagons and diffusons will be approximated by a smooth sigmoid function $\sigma (x)=(1+\mathrm{e}^{-\alpha(x-x_\xi)})^{-1}$, where $\alpha$ and $x_\xi = \hbar \omega_\xi/k_BT$ respectively control the steepness and location of the boundary.  
%at wave number $k = 2\pi/\zeta$, where $\xi$ signifies the critical quantities at the transition and $\zeta$ is the corresponding length scale.} 
A schematic example of this boundary $\sigma$, plotted here as a function of wavevector $k$, is indicated in Fig. \ref{fig1}(b). 
The modes far to the left of the boundary are propagons, while those far to the right are diffusons.  
The modes appearing in the transition region are assigned a mixed character weighted between that of propagons and diffusons.}

Finally, considering together the effects of disorder and the presence of both linear ($\ell$) and parabolic ($p$) modes that appear in low-$D$, the generalized expression for $\kappa$ is  
\begin{equation} \label{eq:kDec}
\kappa = \sum_\ell \kappa_\ell + \sum_p \kappa_p
\end{equation}
where $\kappa_\ell$ and $\kappa_p$ are decomposed into contributions from propagons and diffusons so that  
\begin{widetext}
\begin{eqnarray}
\label{eq:kXg} \kappa_\ell & = &  
\begin{dcases}
\frac{s_d}{d}
\frac{k_B}{(2\pi)^d} \left(\frac{k_B T}{\hbar} \right)^d 
v_\ell^{2-d} \int_0^{x_\ell}{(1-\sigma(x)) \; \tau(x,T) \frac{x^{d+1} e^x}{(e^x-1)^2} \, dx} & \text{(propagons)} \\
 \pi \frac{s_d}{d} \frac{k_B}{(2\pi)^d} \left(k_B \frac{T}{\hbar} \right)^{d-1} v_\ell^{2-d} \int_{0}^{x_\ell}{\sigma(x) \frac{ x^d e^x }{ (e^x-1)^2 } \,dx} & \text{(diffusons)} \hspace{1em},
\end{dcases}  \\ 
\label{eq:kZg} \kappa_p & = & 
\begin{dcases} 
2 \frac{s_d}{d} \frac{k_B}{(2\pi)^d } \left( \frac{k_B T}{\hbar} \right)^{\frac{d+2}{2}} a^{\frac{2-d}{2}} \int_0^{\zRevisions{x_p}} (1-\sigma(x)) \tau(x,T)  \frac{x^{\frac{4+d}{2}} e^x}{(e^x-1)^2} \, dx  & \text{(propagons)} \\
2 \pi \frac{s_d}{d} \frac{k_B}{(2\pi)^d } \left( \frac{k_B T}{\hbar} \right)^{\frac{d}{2}} a^{\frac{2-d}{2}} \int_0^{\zRevisions{x_p}} \sigma(x) \frac{x^{\frac{d+2}{2}} e^x}{(e^x-1)^2} \, dx  & \text{(diffusons)} \hspace{1em},
\end{dcases}
\end{eqnarray}
\end{widetext}
where $x_\ell$ and $x_p$ are the cutoff $x_c$ for the linear $\ell$ and parabolic $p$ modes respectively. 
For the propagons we leave the scattering time $\tau(x,T)$ as--of--yet undetermined; for the diffusons we have used  Cahill's mean free path for the random walk step $\Lambda = \pi v /\omega$. 
Equations (\ref{eq:kDec},\ref{eq:kXg},\ref{eq:kZg}) are the governing equations that we will make use of in the following sections.  
The problem of accurately estimating $\kappa$ is thus reduced to finding a good description of the boundary $\sigma(x)$ and the propagon scattering time $\tau(x)$. 
This decomposition of carriers into propagons and diffusons avoids the previous divergence of the Cahill model, because the lowest frequency modes now retain their propagon character (rather than being assigned a diffuson random walk step $\Lambda=\pi v/\omega$). 

%Before proceeding, we emphasize the three main assumptions that are implicitly incorporated in the circuit model developed here. i) Defects scatter phonons effectively, thereby we can model interface transfer as a diffusion process. For example, our earlier work on 2D graphene/boron nitride superlattices \cite{tzee} show the presence of ``temperature jumps" localized sharply at the interfaces. Similar jumps are observed in Fig. \ref{fig1}(c), which depicts the thermal profile for a diamond nanotube \cite{xxx} sandwiched between a hot and a cold thermal bath; here the jumps occur at the sites of Stone Wales defects. ii) Fine defect modes are not considered, and defect resistances simply add up when interfaces meet. This assumption is rough, but can be ameliorated by the fact that contribution from diffusion (diffusons) can be incorporated into disordered sections, and also that localized modes (locons) contribute negligibly to overall conductivity. iii) Local transport can be diffusive and ballistic, but in this work we study mainly local ballsiticity due to short order range.

\section{Validation and Predictions}

%For lower-dimensional materials such as graphene and carbon nanotubes, we will show that parabolic dispersion affects the temperature dependence of $\kappa$, and disorder can have further consequences. 
%We have chosen these particular materials because of the extensive availability of experimental and/or numerical results, for comparison purposes.

\subsection{Three-dimensional a-SiO$_2$: entire temperature range}

\begin{figure}[!hbtp]
\centering
\includegraphics[width=8cm]{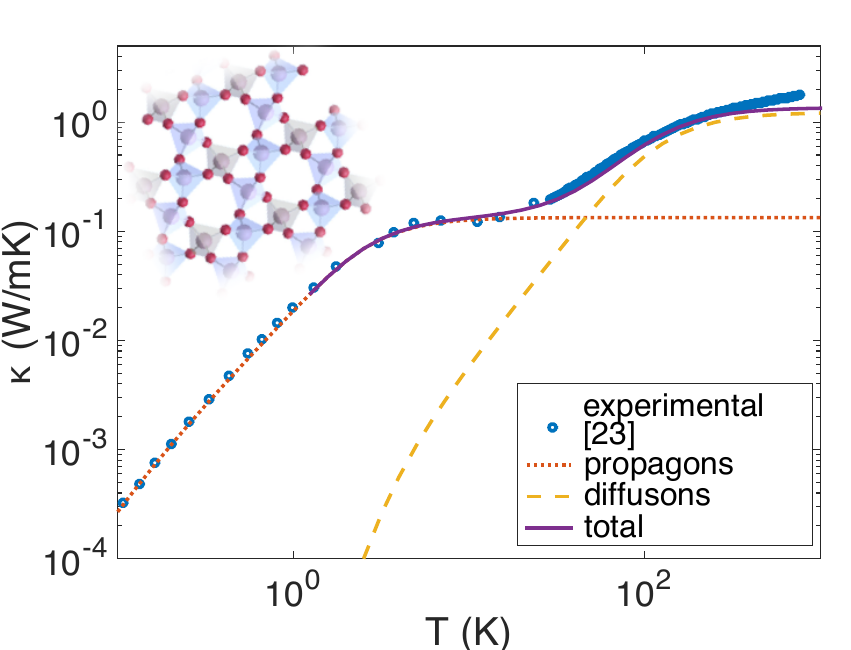}
\caption{\label{fig2} Comparison of generalized model (solid purple line) to experimental results (blue dots) of $\kappa$ {\it vs.} $T$ for 3D amorphous silica.  The isolated contributions of propagons (short-dashed red line) and diffusons (long-broken yellow line) are also shown. The inset schematically illustrates an amorphous silica sample. }
\end{figure}

For 3D amorphous materials it has historically been challenging to capture the low-temperature dependence of $\kappa$  \cite{Cahill1989HeatGlasses,Holland1963AnalysisConductivity}. 
We first validate the disorder model in 3D amorphous materials by comparing it to actual measurements of amorphous silica  reported in Ref. [\onlinecite{Cahill1989HeatGlasses}].  
For a 3D material, there are three acoustic branches (LA, TA$_1$, TA$_2$) exhibiting the usual linear dispersion $\ell$ and there is no contribution from $p$ modes. Equations (\ref{eq:kDec},\ref{eq:kXg},\ref{eq:kZg}) become 
\begin{equation}
\kappa=\sum_{\ell=LA,TA_1,TA_2}{\kappa_\ell}
\end{equation} 
where
\begin{widetext}
\begin{equation} \label{eq:kXg3}
\kappa_\ell  =   
\begin{dcases}
\frac{k_B}{6 \pi^2} \left(\frac{k_B T}{\hbar} \right)^3 
\frac{1}{v_\ell} \int_0^{x_\ell}{(1-\sigma(x)) \; \tau(x,T) \frac{x^{4} e^x}{(e^x-1)^2} \, dx} & \zRevisions{\text{(propagons, 3D)}} \\
\frac{k_B}{6 \pi} \left(\frac{k_B T}{\hbar} \right)^{2} \frac{1}{v_\ell} \int_{0}^{x_\ell}{\sigma(x) \frac{ x^3 e^x }{ (e^x-1)^2 } \,dx}  & \zRevisions{\text{(diffusons, 3D)}} \hspace{1em}.
\end{dcases}
\end{equation}
\end{widetext}
The parameters to be determined are the scattering time $\tau(x)$ and those of the function $\sigma(x)$ that define the propagon/diffuson boundary. 
For scattering time $\tau(x)$, we use common models of boundary scattering $\tau_B$ and defect scattering $\tau_D$. 
Phonon-phonon scattering is neglected as it is small in the temperature range of interest ($T = 0 \, \rm{K}$ to $T = 300 \, \rm{K}$). 
Boundary scattering is given by $\tau_B^{-1}= v_b (1-p_s)/L (1+p_s)$ with $v_b=3/(1/v_l+2/v_t)$ and surface specularity $p_s=0$. 
The specimen length $L=300\,\mu\rm{m}$ as reported in Ref. [\onlinecite{Cahill1989HeatGlasses}], and transverse and longitudinal sound speeds are $v_t=3740$ m/s and $v_l=5980$ m/s respectively \cite{Cahill1989HeatGlasses}. 
For defect scattering\cite{Callaway1959ModelTemperatures,Holland1963AnalysisConductivity,Klemens1958ThermalModes} we use Klemens' scaling relationship $\tau_D \sim \omega^{n-d}$ with $n=3/2$ as suggested by experiment \cite{Klemens1955TheImperfections}, so $\tau_D^{-1} = A x^{3/2}T^{3/2}$, and the factor $A$ is an adjustable parameter. 
The total scattering $\tau(x)$ is then given by Matthiessen's rule $\tau^{-1} = \tau_B^{-1} + \tau_D^{-1}$.
 
The only free parameters in our model are $A$ and those of the function $\sigma(x)=(1+\mathrm{e}^{-\alpha(x-x_\xi)})^{-1}$ to set the propagon/diffuson boundary location and width.  
Here we choose $\alpha \rightarrow \infty$ so that the diffuson/propagon boundary becomes a sharp step function located at frequency $\omega_\xi$, which is also an adjustable parameter, but gives insight to the frequency at which the transition between diffusons/propagons occurs. 
The results are shown in Fig. \ref{fig2}. 
We obtain the best fit to experimental data for $A = 7.4 \times 10^7 \; (k_B/\hbar)^{3/2}$ and $\omega_\xi = 0.25 \,\rm{THz}$.
Our estimate of the propagon/diffuson boundary is not far from the $\sim$ 1THz estimate obtained from molecular dynamics using a modified van Beest potential for amorphous silica\cite{Taraskin2000Ioffe-RegelSilica}.  
Using these parameters, the predictions of the model agree very well with experimental data within the whole temperature range, and ``the plateau'' appears to be the transition regime from propagon-dominated to diffuson-dominated transport.
Diffusons gain dominance as contributors to $\kappa$ as $T$ increases, due to both the increased population of high frequency carriers and the increased scattering of long-wavelength propagons. 

The isolated contributions from diffusons and propagons are also shown in Fig. \ref{fig2}. 
It is evident that the original Cahill minimum thermal conductivity approach captures well the contribution from diffusons which dominate at higher $T$.  
Now, the Callaway contribution of long-wavelength propagons is able to reproduce the low $T$ behavior. 
As a result, the sum of these two contributions matches the experimental results in the whole temperature regime. 
It is interesting to note that the boundary $\omega_\xi$  controls the turning point before the plateau, and the system characteristic length $L$ determines the ultra-low temperature conductivities. 

\subsection{Two-dimensional graphene}

For 2D materials, we consider graphene-like materials, ranging from ordered crystalline to mildly disordered to fully amorphous. 
For a 2D material, Eqs. (\ref{eq:kDec},\ref{eq:kXg},\ref{eq:kZg}) become 
\begin{equation}
\kappa=\sum_{\ell=LA,TA}{\kappa_\ell}+\sum_{p=ZA}{\kappa_p}
\end{equation} 
to reflect that $\kappa$ is the sum of two in-plane linear modes $\ell=(LA,TA)$ and one out-of-plane parabolic mode $p=ZA$. Here,  
\begin{widetext}
\begin{eqnarray}
\label{eq:kXg2D} \kappa_\ell & = &  
\begin{dcases}
\frac{k_B}{4 \pi} \left(\frac{k_B T}{\hbar} \right)^2  
\int_0^{x_\ell}{(1-\sigma(x)) \; \tau(x,T) \frac{x^{3} e^x}{(e^x-1)^2} \, dx} & \zRevisions{\text{(propagons, 2D)}} \\
 \frac{k_B}{4} \left( \frac{k_B T}{\hbar} \right) \int_{0}^{x_\ell} {\sigma(x) \frac{ x^2 e^x }{ (e^x-1)^2 } \,dx} & \zRevisions{\text{(diffusons, 2D)}} \hspace{1em}, 
\end{dcases}  \\ 
\label{eq:kZg2D} \kappa_p & = & 
\begin{dcases} 
\frac{k_B}{2 \pi} \left( \frac{k_B T}{\hbar} \right)^{2} a \int_0^{\zRevisions{x_p}} (1-\sigma(x)) \tau(x,T) \frac{x^{3} e^x}{(e^x-1)^2} \, dx  & \zRevisions{\text{(propagons, 2D)}} \\
\frac{k_B}{2} \left( \frac{k_B T}{\hbar} \right) a \int_0^{\zRevisions{x_p}} \sigma(x) \frac{x^{2} e^x}{(e^x-1)^2} \, dx  & \zRevisions{\text{(diffusons, 2D)}}  \hspace{1em},
\end{dcases}
\end{eqnarray}
\end{widetext}
in which the parameters to be determined are the scattering $\tau(x)$ time and those of the function $\sigma(x)$. 
For the in-plane modes the group velocities are $v_t=13.6 \, \mathrm{km/s}, v_l=21.3 \, \mathrm{km/s}$, for transverse, longitudinal (respectively) and for the parabolic ZA modes the parameter $a=6.2 \times 10^{-7} \, \mathrm{m}^2/\mathrm{s}$.\cite{Pop2012ThermalApplications}

\subsubsection{Description of Scattering}

\begin{table*}[h]
\caption{Scattering models and parameters adopted in this work for linear ($\ell$) and parabolic ($p$) phonon modes. 
These models are extensively used in single mode relaxation time modeling\cite{Verma2013AGraphene,Peierls1955QuantumSolids,Klemens1958ThermalModes, Morelli2002EstimationSemiconductors,Balandin2008SuperiorGraphene}.
The atomic mass of a carbon atom C is $M=1.99 \times 10^{-26} \, \mathrm{kg}$. $S_0=\delta a_0$ is the cross section of each C atom, with $\delta=3.35 \, \mathrm{\AA}$ as graphene thickness, and $a_0=1.42 \, \mathrm{\AA}$ the length of a C-C bond. 
The volume per C atom is $V_m=\delta/n$ where $n=3.82 \times 10^{19} \, \mathrm{m}^{–2}$ is the number of atoms per unit area. 
$\Gamma_m=7.54 \times 10^{-5}$ represents the defect scattering in pristine graphene based on the natural isotopic abundance of $98.9\% \; ^{12}$C and $1.1\% \; ^{13}$C\cite{Lindsay2010FlexuralGraphene}. 
The Gruneisen parameters are $\gamma_{\ell}=2$ for linear modes \cite{Klemens1994ThermalPlane}, and $\gamma_p=-18.64$ for parabolic modes\cite{Kong2009First-principlesGraphene}. 
$\Theta$ is the Debye temperature.}
\label{tab:scat}
\begin{tabular}{l l l }
\hline \hline
Scattering Mechanism & scattering model   & parameters \\\hline 
Boundary scattering &  
$\begin{array} {lcl} \tau_{B,\ell}^{-1}&=&v_\ell/L \\ \tau_{B,p}^{-1}&=&v_p/L=A_B x^{1/2} T^{1/2} \end{array}$ &
$ A_B=2(a k_B/\hbar)^{1/2}/L $   \\
\hline
Defect scattering (Rayleigh) &  
$\begin{array} {lcl} \tau_{I,\ell}^{-1}&=&A_{I,\ell} x^3 T^3  \\
\tau_{I,p}^{-1}&=&A_{I,p} x^2 T^2  \end{array}$  &
$\begin{array} {rl} A_{I,\ell}&= S_0 \Gamma_m (k_B/ \hbar)^3 /4 v^2    \\
A_{I,p}&=S_0 \Gamma_m (k_B/ \hbar)^2 /8 a \end{array}$   \\
\hline
Umklapp process &  
$\begin{array} {lcl} \tau_{U,\ell}^{-1}&=&A_{U,\ell} x^2 T^3 \; \mathrm{e}^{-\Theta_\ell/3T} \\
\tau_{U,p}^{-1}&=&A_{U,p} x^{-2}  \; \mathrm{e}^{-\Theta_p/3T} \end{array}$ &
$\begin{array} {rl} A_{U,\ell}&=k_B^2 \gamma_\ell^2 /\hbar M v_\ell^2\Theta_\ell \\
A_{U,p}&=2\gamma_p^4 \hbar^2 \omega_B/ 27 M^2 a^2 \\
\omega_B&=28 \mathrm{GHz} \cite{Verma2013AGraphene} \end{array}$   \\
\hline
Normal process &  
$\begin{array} {lcl} \tau_{N,LA}^{-1}&=&A_{N,LA} x^2 T^5 \\
\tau_{N,TA}^{-1}&=&A_{N,TA} x T^5 \end{array}$ &
$\begin{array} {rl} A_{N,\ell}&=k_B^5 \gamma_\ell^2 V_m/\hbar^4 M v_\ell^5    \end{array}$   \\
\hline \hline
\end{tabular}
\end{table*}

To utilize Eqs.~(\ref{eq:kXg2D},\ref{eq:kZg2D}) scattering models for $\tau(x)$ for in-plane and out-of-plane modes need to be selected. 
The descriptions adopted here are summarized in Table \ref{tab:scat}. 
For all forms of scattering, we differentiate between linear and parabolic modes; the latter are discussed in detail in Ref. [\onlinecite{Mingo2005LengthWaves.}].
Unless otherwise stated, in all cases the parameters in Table \ref{tab:scat} are obtained directly from experimental or density functional theory (DFT) results ({\it i.e.}, no parameters are fitted). 
From Table \ref{tab:scat}, at low temperatures boundary scattering is dominant but as temperature increases, first defect scattering and then inter-phonon scattering will successively become dominant. 
We employ the Peierls-Klemens model to represent the first-order Umklapp processes. 
%Although they do not contribute to phonon viscosity, N processes still redistribute phonon populations and affect $\kappa$ indirectly. 
As suggested in Ref. [\onlinecite{Mingo2005LengthWaves.}], N processes for ZA modes are assumed to be higher-order effects and are not considered. 

\subsubsection{Crystalline Graphene}

\begin{figure*}[!hbtp]
\centering
\includegraphics[width=0.9\textwidth]{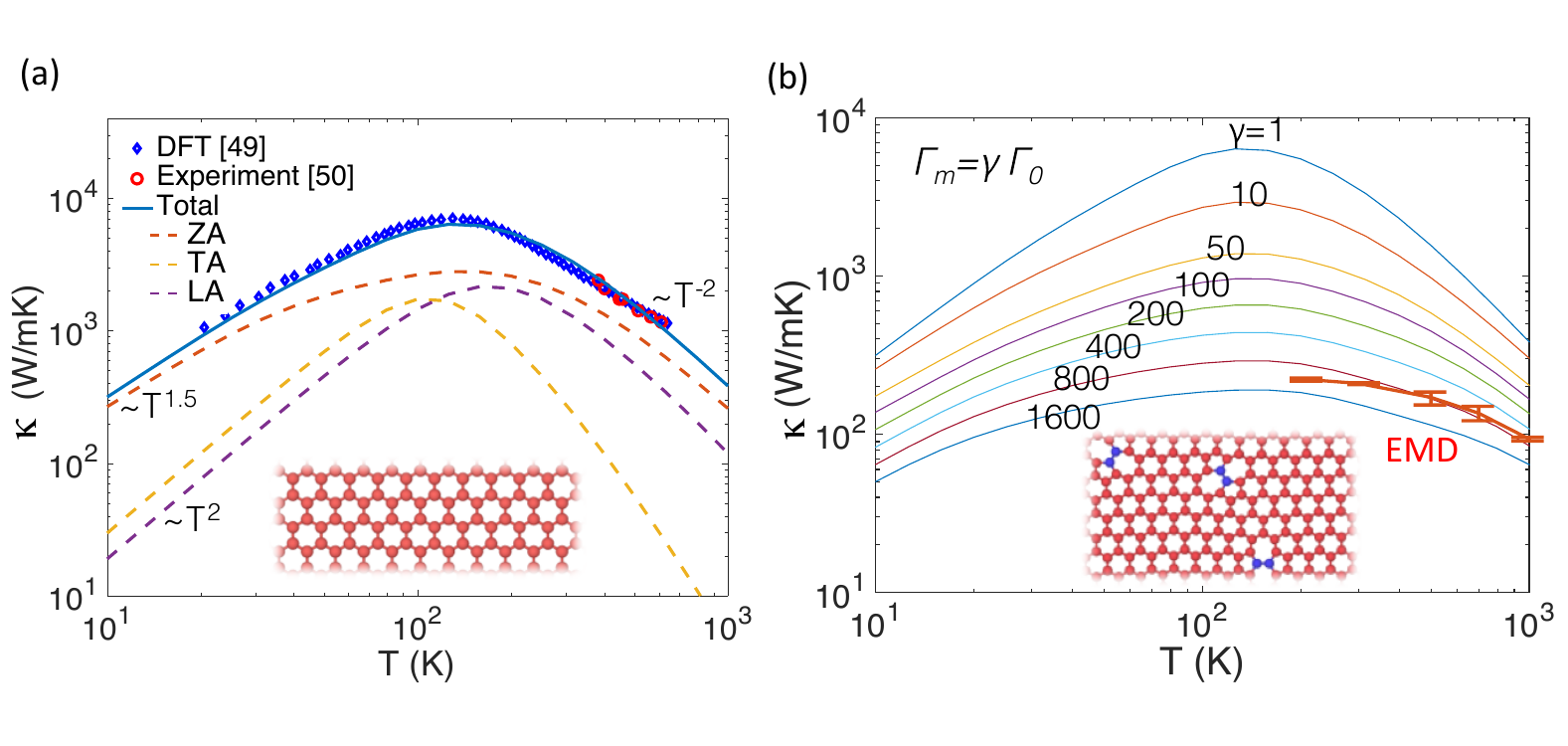}
\caption{\label{fig3} (a) Comparison of generalized model to available experimental and density functional theory results of $\kappa$ {\it vs.} $T$ for pristine graphene. 
The generalized model predicts a $T^{3/2}$ scaling at low $T$ consistent with a predominant contribution from ZA modes, and a $T^{-2}$ scaling at high temperature.
Individual contributions from ZA, TA, and LA modes are also shown. 
(b) Influence of Stone-Wales defects on $\kappa$ {\it vs.} $T$ according to both equilibrium molecular dynamics, and  generalized model for different degrees of defect scattering $\Gamma_m = \gamma \Gamma_O$ where $\Gamma_O$ is the defect scattering for crystalline graphene (see Table II). 
The statistical error bars marked on the equilibrium molecular dynamics results correspond to the first standard deviation.
} 
\end{figure*}

The thermal transport properties of even nominally crystalline graphene continue to be the subject of research attention. 
Accurate experimental and computational studies have been published recently \cite{Balandin2011ThermalMaterials, Kong2009First-principlesGraphene, Lindsay2010FlexuralGraphene}, but there is still debate about the nature of the dominant carrier modes. 
On one hand, first-principles calculations show that ZA modes contribute the most ($76\%$) to the thermal conductivity at room temperature\cite{Lindsay2010FlexuralGraphene}. 
Similarly, it has been shown that by including only ZA modes but neglecting all others, a Callaway approach with an appropriate relaxation model \cite{Mingo2005LengthWaves.} can accurately account for $\kappa$ within the whole temperature range \cite{Mariani2008FlexuralGraphene}. 
On the other hand, it has also been suggested that ZA modes contribute negligibly to overall thermal conductivity, due to their low group velocities but large Gruneisen parameters.\cite{Balandin2011ThermalMaterials, Balandin2008SuperiorGraphene, Kong2009First-principlesGraphene}   

To provide some insights, we consider the predictions of our generalized model for crystalline graphene.  
For the crystalline case, there are only propagons so $\sigma(x) = 0$. 
Using the descriptions of scattering in Table \ref{tab:scat}, our results in comparison to both DFT  \cite{Lindsay2014PhononPrinciples} and experimental \cite{Chen2011RamanEnvironments} results are shown in Fig. \ref{fig3}(a). 
We obtain a close match to the DFT calculation (blue diamonds) \cite{Lindsay2014PhononPrinciples} throughout the entire temperature regime, with no adjustable parameters. 
In the high temperature regime, our results match both the DFT results and the available experimental results, but the predicted scaling appears to be more similar to the experimental measurements. 

Apart from the good agreement, there are several observations for the low, intermediate, and high temperature regime.
From Eqs.~(\ref{eq:kXg2D},\ref{eq:kZg2D}) and Table \ref{tab:scat}, the temperature dependence of $\kappa$ for linear and parabolic modes depends on the dominant scattering mechanism. 
(i) At low-temperatures when boundary scattering is dominant, $\kappa_\ell \sim T^2$ and $\kappa_p \sim T^{3/2}$. 
Our analysis predicts a $T^{3/2}$ dependence, and thus a dominant contribution of ZA modes. 
(ii) As temperature increases to an intermediate regime, the scaling may change for several reasons. 
More LA/TA modes are excited and their influence can change the temperature dependence. 
Additionally, other forms of scattering (defect and/or phonon-phonon) may emerge and also change the scaling.
(iii) In the high temperature regime when phonon-phonon scattering is dominant, our predictions contain some ambiguity due to the uncertainty of the scattering model for ZA modes. 
However, the high-order scattering model used here successfully captures the $T^{-2}$ behavior at high temperature measured from experiments\cite{Chen2011RamanEnvironments} and continuum-theoretical predictions \cite{munoz2010ballistic}. 
This is in contrast to the DFT results which instead predict a $T^{-1.5}$ dependence. 

\subsubsection{Crystalline Graphene with Stone Wales Defects}

Although impedance of phonon conduction due to the presence of vacancies has been studied \cite{Hao2011MechanicalDefects}, the detailed temperature dependence of graphene with defects has not yet been established. 
We consider here how a mild distribution of Stone-Wales defects affects $\kappa$. 
Since no experimental results are available, we use equilibrium molecular dynamics (EMD) and the Green-Kubo formulation to calculate $\kappa$ and compare to the results of our model. 
The optimized Tersoff potential is used, and simulation details are available in Ref. [\onlinecite{ZhuErtekinNL}]. 
The sample size is $1.25\times 2.16$ nm$^2$, which is $50\times 50$ unit cells.
The sample, shown in Fig. \ref{fig3}(b), is produced by selecting bonds at random and rotating by ninety degrees (Stone-Wales transformation). 
The system is then relaxed before increasing its temperature. 
The defect density in Fig. \ref{fig3}(b) corresponds roughly to 0.38 defects per nm$^2$ (equivalent to $8\%$ defective unit cells). 

In our model, we consider the material as crystalline, but incorporate the effects of the Stone-Wales defects indirectly through the scattering parameter $\Gamma_m=\gamma \Gamma_0$, where $\Gamma_0=7.54 \times 10^{-5}$ is the value for natural graphene due to its isotopic composition (see Table \ref{tab:scat}). 
As shown in Fig. \ref{fig3}(b), the Stone-Wales defects reduce $\kappa$ as well as its temperature sensitivity. 
The EMD results are in reasonable agreement with the generalized model at high temperatures for the choice $\gamma \approx 800$.

\subsubsection{Amorphous Graphene}

\begin{figure}[!hbtp]
\centering
\includegraphics[width=8.5cm]{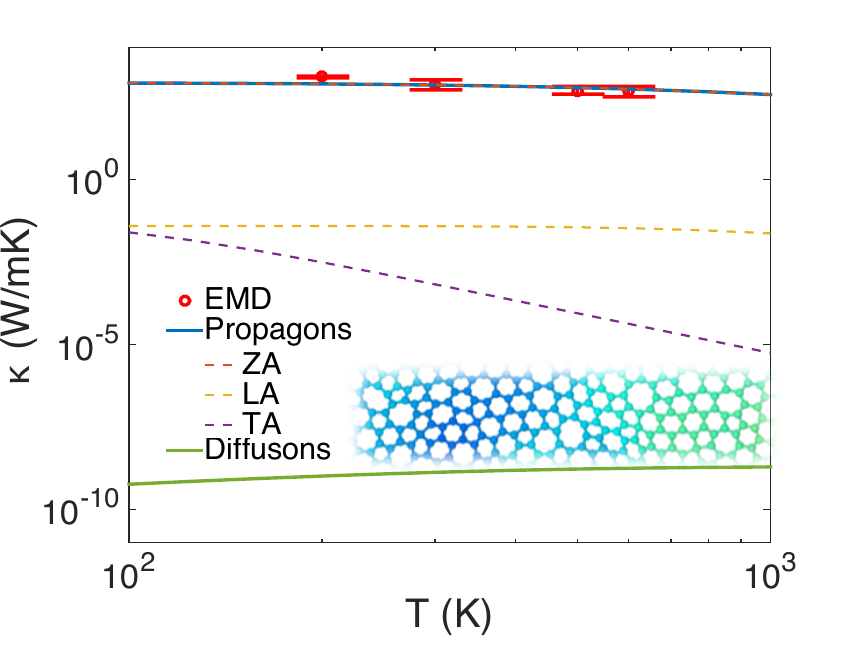}
\caption{\label{fig4} $\kappa$ {\it vs.} $T$ according to generalized model and equilibrium molecular dynamics for amorphous graphene. 
The dominant carriers are predicted to be out-of-plane (ZA) propagons. 
Unlike 3D amorphous materials diffusons contribute negligibly to $\kappa$ over the entire temperature range.
The statistical error bars marked on the equilibrium molecular dynamics results correspond to the first standard deviation.}
\end{figure}

We now consider ``amorphous graphene", a 2D sheet that maintains the $sp^2$ bond order of pristine graphene, but contains a disordered distribution of rings of different sizes varying from 4-8 atoms \cite{Kotakoski2011FromCarbon}. 
%Nanoporous graphene has also recently been synthesized, and proposed for use in water desaliation\cite{Surwade2015WaterGraphene.}
For amorphous graphene, both diffusons and propagons are present, and now contributions from all parts of Eqs. (\ref{eq:kXg2D},\ref{eq:kZg2D}) give rise to the total 
	$\kappa$. 
Similar to 3D amorphous silica case, the Ioffe-Regel boundary between diffusons and propagons is of importance.  
We compare the results of the generalized model to our EMD simulation results \cite{ZhuErtekinNL}.

The sample is generated following the procedure outlined in Ref. [\onlinecite{Kumar2012AmorphousGlass}]: pristine samples of the same size are first melted into 2D carbon gases at $T = 4500$ K, then quenched to the target temperature in 1 ns, which is followed by a Nose-Hoover thermostating for 0.5 ns. 
The amorphous graphene buckles naturally, as shown in Fig. \ref{fig4}, where the buckling height is denoted by the color map. 
Note that the current samples are homogeneously {\it sp}$^2$-bonded carbon materials, and thus different from those generated by introducing vacancies \cite{Carpenter2012AnalysisGraphene} where dangling carbon bonds are present.

The thermal conductivity of amorphous graphene, obtained both from EMD and the generalized model, is plotted in Fig. \ref{fig4}. 
This thermal conductivity is suppressed by a factor of 1.65 at 300K in comparison to crystalline graphene, for an equivalent sized system. 
For the generalized model, we have again assumed a sharp diffuson/propagon boundary $\omega_\xi$ and fitted it to best match the EMD results.  
The best match corresponds to $\omega_\xi = $ 0.8 THz, which gives very reasonable agreement with the EMD results. 
Remarkably, this also agrees very well with our estimate from phonon localization analysis in which the boundary is obtained from phonon modal diffusivities   \cite{ZhuErtekinNL}. 
It is encouraging that two independent approaches yield a very similar estimate of the boundary. 

There are some interesting differences to note in the predicted trends for 2D amorphous systems, in comparison to their 3D counterparts.  
There is no ``plateau region", nor is there an observable transition from propagon to diffuson -dominated transport.  
In fact, Fig. \ref{fig4} also shows the separate contributions of the propagons and the diffusons, from which it is evident that diffusons barely contribute to the overall $\kappa$ up to temperatures as large as $T = $ 1000K. 
As described in detail in Ref. [\onlinecite{ZhuErtekinNL}], we speculate that this arises from the inherent difference in the nature of random walks in different dimensions: random walks of dimension $d=1,2$ are recurrent, while those of dimension $d=3$ are transient. 
Moreover, it is noteworthy that throughout the entire temperature range, the generalized model predicts that the out-of-plane ZA modes dominate the heat transport for the amorphous system.

\subsection{One-dimensional and quasi one-dimensional nanotubes and nanothreads}

\begin{figure}[!hbtp]
\centering
\includegraphics[width=8.5cm]{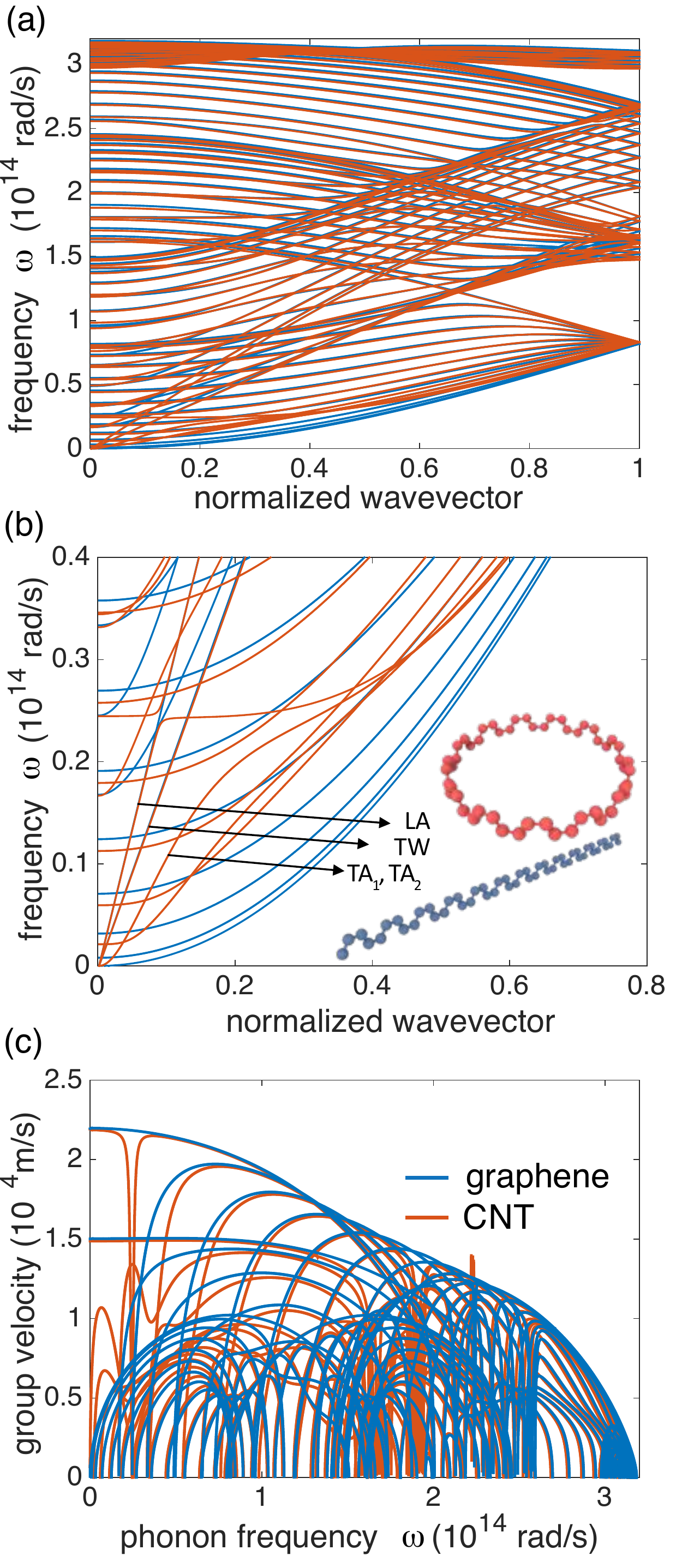}
\caption{\label{fig5} 
\zRevisions{(a) Phonon dispersion of a (13,13) CNT (red) and of zone-folded graphene (blue). 
(b) Zoomed-in plot of the dispersions near the Brillouin zone $\Gamma$ point; the rolled tube LA, TW, and TA$_1$, TA$_2$ modes are marked by arrows. The inset shows the unit cells used for the (13,13) CNT and for zone-folded graphene. 
(c) Corresponding group velocities for both cases. }} 
\end{figure} 

For our analysis of ordered and disordered 1D systems, we consider carbon nanotubes and disordered diamond nanothreads. 
Phonon transport on carbon nanotubes (CNTs) is featured by its sensitivity to the tube  radius. \cite{Saito1998PhysicalNanotubes,Marconnet2013ThermalMaterials,Yue2015DiameterInitio} 
To a first approximation, the CNT phonon dispersion can be considered to be a zone-folded dispersion of 2D graphene \cite{Saito1998PhysicalNanotubes}. 
This approximates most modes well, but is less accurate for the low-energy phonons. 
When rolled into a tube, the graphene LA modes remain effectively unchanged, but the graphene TA modes become the nanotube twist (TW) modes, the graphene down-to-zero flexural ZA modes transform into non-zero breathing modes, and a new set of TA modes (TA$_1$, TA$_2$) unique to the rolled system emerges. 
The latter two considerations cause discrepancies between the actual dispersion of a carbon nanotube, and the equivalent zone-folded graphene dispersion. 

For example, Fig. \ref{fig5} shows a comparison of the actual dispersion of a (13,13) CNT (red lines) to that of appropriately zone-folded graphene (blue lines), obtained by direct solution of the eigen-problem of the dynamical matrix.
We use the (13,13) CNT here, since its radius is close to the one for which $\kappa$ has been measured in experiments \cite{Pop2006ThermalTemperature} to which we will compare. 
Of the down-to-zero modes, the LA and the TW modes clearly exhibit an acoustic nature, while the degenerate TA$_1$,TA$_2$ modes exhibit a more quadratic nature.  
For the latter set, the transition from parabolic to linear dispersion only becomes complete in the limit of vanishing radius (truly 1D systems); the (13,13) CNT dispersion shown in Fig. \ref{fig5} therefore shows remnants of 2D dispersion and in some sense this CNT can be considered a quasi-1D system.  

An interesting question is ``for which diameter $D$ will the CNT thermal properties be close to that of a true 1D system?". 
We assume that the CNT dispersion will reduce to that of graphene when $D \gg \lambda_0$, where $\lambda_0$ represents a dominant graphene phonon wavelength. 
\zRevisions{We define the ballistic transporting capability $K(x)$ of a parabolic mode with $x = \hbar\omega/k_BT$ from Eq. (\ref{br}) with $d=2$ as}  
\begin{eqnarray}
\kappa_p & = & \frac{k_B}{4\pi} \left( \frac{k_B T}{\hbar} \right)^{\frac{3}{2}} a^{-\frac{1}{2}} \int_0^{x_p} \frac{\Lambda(x,T) x^{\frac{5}{2}} e^x}{(e^x-1)^2} \, dx \nonumber \\ 
  & = & \frac{k_B}{4\pi} \left( \frac{k_B T}{\hbar} \right)^{\frac{3}{2}} a^{-\frac{1}{2}} \int_0^{x_p} \Lambda(x,T) K(x) \, dx \hspace{1em},  
\end{eqnarray} 
\zRevisions{so that}
\begin{equation}
K(x)=\frac{x^{5/2} e^x}{(e^x-1)^2} \hspace{1em}. 
\end{equation} 
Then $K(x)$ is maximized for $x_0 \approx 1.77603$, or 
\begin{equation}
\lambda_0= 2 \pi (\frac{k_B T}{\hbar a}x_0)^{-1/2}	\hspace{1em}. 
\end{equation} 
For graphene, $\lambda_0 \approx 10.326/\sqrt{T}$ nm, which sets the critical diameter to $D = 6.0$ {\AA} at $T = $ 300 K. 
Therefore, for most experimental data, where usually $D>1$ nm, phonon transport may resemble almost 2D transport.
For the (13,13) CNT pictured in Fig. \ref{fig5} the diameter is close to 18 {\AA} and the remnant parabolic dispersion is clear for the modes labeled TA$_1$,TA$_2$. 

\subsubsection{(13,13) Carbon Nanotube}

\begin{figure*}[!hbtp]
\centering
\includegraphics[width=1\textwidth]{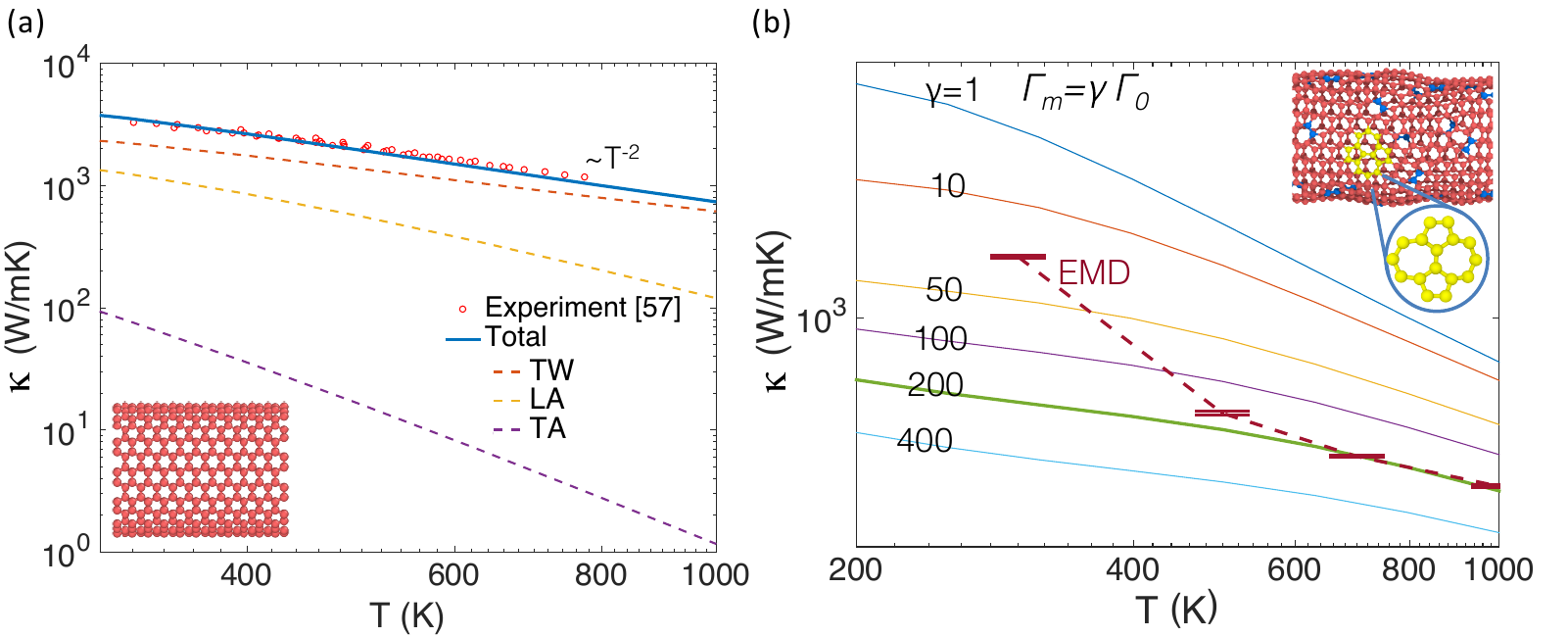}
\caption{\label{fig6} (a) $\kappa$ {\it vs.} $T$ for (13,13) CNT according to generalized model, compared to the experimental results measured for a nanotube of similar radius. 
(b) The influence of Stone-Wales defects on $\kappa$ {\it vs.} $T$, according to equilibrium molecular dynamics and generalized model.  
For the generalized model the influence of defects is accounted for by the defect scattering $\Gamma_m = \gamma \Gamma_0$, where $\Gamma_0$ is the defect scattering parameter for the pristine material (see Table II). 
The statistical error bars marked on the equilibrium molecular dynamics results correspond to the first standard deviation.} 	
\end{figure*}

Based on the discussion above we model the (13,13) CNT thermal conductivity as 
\begin{equation}
\kappa=\sum_{\ell=LA,TW}{\kappa_\ell} +\sum_{p=TA_1,TA_2}{\kappa_p} 
\hspace{1em},
\end{equation}
where $\kappa_\ell$ and $\kappa_p$ are given by the 2D description in Eqs. (\ref{eq:kXg2D}) and (\ref{eq:kZg2D}).
%\begin{widetext}
%\begin{eqnarray}
%\label{eq:k1Dl} \kappa_\ell & = &  
%\frac{s_d}{d}
%\frac{k_B}{(2\pi)^d} \left(\frac{k_B T}{\hbar} \right)^d 
%v_\ell^{2-d} \int_0^{x_\ell}{\tau(x) \frac{x^{d+1} e^x}{(e^x-1)^2} \, dx}   \\ 
%\label{eq:k1Dp} \kappa_p & = & 
%2 \frac{s_d}{d} \frac{k_B}{(2\pi)^d } \left( \frac{k_B T}{\hbar} \right)^{\frac{d+2}{2}} a^{\frac{2-d}{2}} \int_0^{x_c} \tau(x)  \frac{x^{\frac{4+d}{2}} e^x}{(e^x-1)^2} \, dx  \hspace{3em}  
%\end{eqnarray}
%\end{widetext}
Only propagon contributions are included for the ordered system, and we use the same parameters as for graphene in the previous section, except that the ZA mode disappears, and the TA$_1$, TA$_2$ mode group velocities from the dispersions in Fig. \ref{fig5} are both 9.4 km/s.
 The description of scattering in Table \ref{tab:scat} is used again. 
%As shown in Fig. \ref{fig5}, LA and TW modes are barely modified by rolling. 
%Applying the similar relaxation models as graphene, 
This approach is able to reproduce the available experimental data (see Fig. \ref{fig6}(a)) \cite{Pop2006ThermalTemperature}, also with no adjustable parameters. 
The dominant contribution to $\kappa$ throughout the full temperature range comes from the twist mode TW. 
As discussed in Ref. [\onlinecite{Pop2006ThermalTemperature}], the measured temperature dependence of $\kappa$ arises from a  competition between three-phonon scattering processes. Above room temperature it was fitted as $1/(A T+ B T^2)$, where $A, B$ are constants, and is thus dominated by $T^{-2}$ at high temperature. 
The scaling is well-captured by the current model.

The influence of Stone-Wales defects is also shown in Fig. \ref{fig6}(b), for a 4\% defect density at randomly selected sites. 
Fig. \ref{fig6}(b) shows the resulting $\kappa$ according to EMD results as well as the model predictions for difference degrees of defect scattering incorporated through the scattering parameter $\Gamma_M = \gamma \Gamma_O$ (see Table II). 
The defects reduce the thermal conductivity approximately 3-fold compared to the pristine CNT.

\subsubsection{Diamond Nanothreads} 

\begin{figure*}[!hbtp]
\centering
\includegraphics[width=1\textwidth]{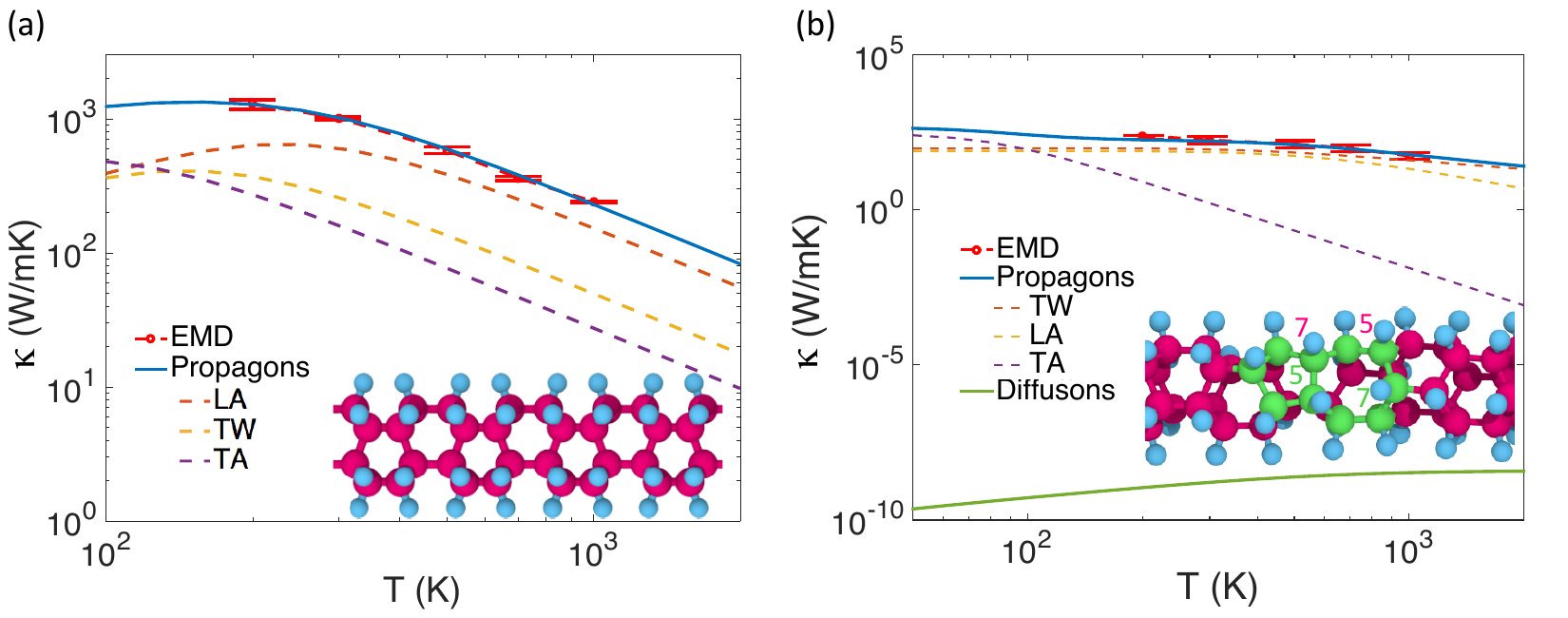}
\caption{\label{fig7} $\kappa$ {\it vs.} $T$ for (a) a  crystalline (3,0) hydrogenated {\it sp}$^3$ carbon nanotube  and (b) a glassy diamond nanothread, both according to equilibrium molecular dynamics and generalized model. 
The generalized model predicts dominant contributions from LA modes for the pristine system; for the glassy case the contributions of propagons is dominant throughout the entire temperature range and the diffuson contribution is negligible.
The statistical error bars marked on the equilibrium molecular dynamics results correspond to the first standard deviation.} 
\end{figure*}

One-dimensional diamond nanothreads have been recently synthesized in the laboratory for the first time\cite{Fitzgibbons2014Benzene-derivedNanothreads}. 
The thermal properties of an actual 1D system, particularly a highly disordered one, may be better exhibited by these nanothreads.   
\zRevisions{Diamond nanothreads are based on (3,0) nanotubes, but differ because they (i) are hydrogenated so that the bonding exhibits an $sp^3$ configuration, and (ii) contain a random distribution of Stone Wales defects at high density ($\approx 20\%$) is present, introducing structural disorder. }
We consider $\kappa$ for both a pristine (3,0) hydrogenated system and a disordered system with 20\% Stone-Wales defects introduced at random sites.  
Since the radius of a (3,0) CNT is only 4 {\AA} we use a true 1D representation.  
All modes are \zRevisions{approximated as linear in wavevector\cite{Saito1998PhysicalNanotubes}}, and the thermal conductivity is given by 
\begin{equation}
\kappa=\sum_{\ell=LA,TA_1,TA_2,TW}{\kappa_\ell} \hspace{1em}, %+\sum_{\zeta=RBM}{\kappa_\zeta}
\end{equation}
where 
\begin{widetext}
\begin{equation}
\label{eq:k1D} \kappa_\ell  =   
\begin{dcases}
\frac{k_B^2 T}{\pi \hbar}
v_\ell \int_0^{x_\ell}{(1-\sigma(x)) \; \tau(x,T) \frac{x^2 e^x}{(e^x-1)^2} \, dx} & \zRevisions{\text{(propagons, 1D)}} \\
 k_B v_\ell \int_{0}^{x_\ell}{\sigma(x) \frac{ x e^x }{ (e^x-1)^2 } \,dx} & \zRevisions{\text{(diffusons, 1D)}} \hspace{1em}. 
\end{dcases} 
\end{equation}
\end{widetext}
We use the same scattering parameters we used for graphene in the previous sections, except that the ZA mode disappears, and the TA and TW group velocities are reduced to 8.1 km/s and 12.4 km/s respectively, as obtained from lattice dynamics.
The thermal conductivity of the 1D ultra-thin nanotube is plotted in Fig. \ref{fig7}(a). 
Compared to the (13,13) nanotubes, the conductivity is reduced by a factor of three, due largely to the reduction of group velocities.

The amorphous version is shown in Fig. \ref{fig7}(b), according to the generalized model and EMD simulations. 
%To study the amorphous limit of the 1D system, we carefully quantify the effects of localization and scattering via disorder, which will be presented in another paper \cite{ZhuErtekinNL}. 
%In this work, we calculated the temperature dependence of the thermal conductivity for diamond nanothreads via EMD and the classical model (See Fig. \ref{fig7}). 
For the generalized model, we have again assumed a sharp diffuson/propagon boundary $\omega_\xi$ and fitted it to best match the EMD results.  
The best match gives $\omega_\xi = $ 0.45 THz. 
This estimate of Ioffe-Regel boundary also agrees very well with our phonon localization analysis \cite{ZhuErtekinNL}. 
In the disordered system $\kappa$ is suppressed by a factor of 5 at $T = $ 300K in comparison to the crystalline (3,0) hydrogenated tube.
The twisting modes are predicted to be the predominant energy carriers.
Similar to 2D amorphous graphene, diffusons are observed to contribute negligibly to overall $\kappa$.
Furthermore, once again in contrast to 3D, there is no ``plateau region" nor is there a corresponding transition from propagon-dominated to diffuson-dominated transport.

\section{conclusion}
We have presented a generalized framework to describe the lattice thermal conductivity of low-dimensional and disordered materials. 
The approach is motivated by the Allen-Feldman description of thermal transport in amorphous 3D materials, in which heat carriers are categorized as propagons and diffusons based on their transporting capacity. 
Results of the model are compared to experimental measurements and/or equilibrium molecular dynamics simulations, and show good agreement. 
Some interesting aspects to thermal transport in low-dimensional and disordered materials are suggested, including a more mild suppression of the thermal conductivity in comparison to 3D, the lack of a ``plateau" in the temperature dependence of the thermal conductivity, and the negligible contribution of diffusons to the transport.

%%%%%%%%%%%%%%%%%%%%%%%%%%%%%%%%%%%%%%%%
%%%%%%%%%%%%%%%%%%%%%%%%%%%%%%%%%%%%%%%%
%%%%%%%%%%%% APPENDIX BELOW %%%%%%%%%%%%%%
%%%%%%%%%%%%%%%%%%%%%%%%%%%%%%%%%%%%%%%%
%%%%%%%%%%%%%%%%%%%%%%%%%%%%%%%%%%%%%%%%

%%%%%%%%%%%%%%%%%%%%%%%%%%%%%%%%%%%%%%%%
%%%%%%%%%%%%%%%%%%%%%%%%%%%%%%%%%%%%%%%%
%%%%%%%%%%%% ACKNOWLEDGEMENTS BELOW %%%%%%%%%%%%%%
%%%%%%%%%%%%%%%%%%%%%%%%%%%%%%%%%%%%%%%%
%%%%%%%%%%%%%%%%%%%%%%%%%%%%%%%%%%%%%%%%

\section*{Acknowledgement}
% \addcontentsline{toc}{section}{Acknowledgement}
We are grateful to David Cahill for insightful discussions on amorphous models. This work is supported by the National Science Foundation through grant no. CBET-1250192.  
We also acknowledge the support from various computational resources: this research is part of the Blue Waters sustained-petascale computing project, which is supported by the National Science Foundation (awards OCI-0725070 and ACI-1238993) and the state of Illinois.  
Blue Waters is a joint effort of the University of Illinois at Urbana-Champaign and its National Center for Supercomputing Applications.  
Additional resources were provided by (i) the Extreme Science and Engineering Discovery Environment (XSEDE) allocation DMR-140007, which is supported by National Science Foundation grant number ACI-1053575, and (ii) the Illinois Campus Computing Cluster.

%%%%%%%%%%%%%%%%%%%%%%%%%%%%%%%%%%%%%%%%
%%%%%%%%%%%%%%%%%%%%%%%%%%%%%%%%%%%%%%%%
%%%%%%%%%%%% BIBLIOGRAPHY %%%%%%%%%%%%%%
%%%%%%%%%%%%%%%%%%%%%%%%%%%%%%%%%%%%%%%%
%%%%%%%%%%%%%%%%%%%%%%%%%%%%%%%%%%%%%%%%

\bibliography{classicK}

\end{document}